\documentclass[12pt]{article}

\usepackage{epsfig}
\usepackage{bm}
\usepackage{ametsoc}

\catcode`\"=\active \let"=\"
\let\3=\ss

\begin{document}

\title{\bf Condensation of cloud microdroplets in homogeneous
  isotropic turbulence} \author{Alessandra S. Lanotte\\ CNR - Istituto
  di Scienze dell'Atmosfera e del Clima, \\ Via Fosso del Cavaliere
  100, I-00133 Roma, Italy \\ and INFN, Sezione di Lecce, I-73100
  Lecce, Italy\\ \and Agnese Seminara\thanks{\textit{Corresponding
      author address:} Agnese Seminara Harvard University, School of
    Engineering and Applied Sciences, 29 Oxford Street, 02138,
    Cambridge - MA, USA. \newline{E-mail:
      seminara@seas.harvard.edu}}\\Harvard University, School of
  Engineering and Applied Sciences, \\ 29 Oxford Street, 02138,
  Cambridge - MA, USA \\ \and Federico Toschi\\ CNR - Istituto per le
  Applicazioni del Calcolo, I-00185 Roma, Italy\\ INFN, Sezione di
  Ferrara, Ferrara, Italy } \amstitle

\begin{abstract}
  The growth by condensation of small water droplets in a
  three-dimensional homogeneous isotropic turbulent flow is
  considered. Within a simple model of advection and condensation, the
  dynamics and growth of millions of droplets are integrated. A
  droplet-size spectra broadening is obtained and it is shown to
  increase with the Reynolds number of turbulence, by means of two
  series of direct numerical simulations at increasing
  resolution. This is a key point towards a proper evaluation of the
  effects of turbulence for condensation in warm clouds, where the
  Reynolds numbers typically achieve huge values. The obtained
  droplet-size spectra broadening as a function of the Reynolds number
  is shown to be consistent with dimensional arguments. A
  generalization of this expectation to Reynolds numbers not
  accessible by DNS is proposed, yielding upper and lower bounds to
  the actual size-spectra broadening.  A further DNS matching the
  large scales of the system suggests consistency of the picture
  drawn, while additional effort is needed to evaluate the impact of
  this effect for condensation in more realistic cloud conditions.

\end{abstract}

\section{Introduction}
\label{sec:intro}
The growth of microdroplets by condensation is a long-standing problem
of cloud physics (\cite{prupp}), meteorology (see e.g. \cite{meteo}),
medicine (\cite{martonen}) and engineering (\cite{engines}). A
fundamental understanding of key issues such as the turbulent mixing
inside clouds, or the interaction of turbulence with microphysics is
important for a variety of applications (the parameterization of small
scales in large scales models, the analysis of radiative transfer
through clouds, the accurate prediction of the initiation of
precipitation). The peculiar features of turbulence, influencing the
motion on a wide range of space-time scales, can alter significantly
the condensation process, providing a strongly fluctuating and
intermittent moist environment. This is a well-known concept in
engineering, where turbulence is considered as a key ingredient for
the optimization of fuel-air mixing and of the rate of fuel
vaporization in engines (\cite{engines}). Similar ideas have a long
history in cloud physics: turbulence plays a role both in collision
processes
(\cite{saffman,shaw,collins,hunt,natureGrisha,wang06,wilkinson}) and
in condensation. In this work we will focus particularly on the latter
process.

Condensation is a fundamental process for the early stages of cloud
evolution. It is the only mechanism providing a growth of cloud droplets
immediately after their formation, when they are few micrometers
in size. The initiation of collisions and coalescence occurs when a
few droplets become large enough to fall - a radius of $20 ~\mu m$ is
commonly considered as a threshold. For these droplets, collisions are much more
efficient than condensation as a growth mechanism. Indeed, after
nucleation, droplet radius grows roughly one order of magnitude by
condensation. In a timescale comparable to that of condensation,
collisions produce raindrops, bridging a gap of about two orders of
magnitude in size. However, the high efficiency of the latter is
strongly influenced by the general features of the former. In
particular, gravitational collisions are highly effective when 
the previous condensation process produces a
population of droplets spanning a large variety
of sizes. Uniform condensational growth leads to narrow droplet-size
spectra instead (see e.g. \cite{squires52,leeprupp77}). Indeed,
provided that all droplets grow in similar ambient conditions, small
droplets grow faster than large ones and thus all droplets finally
tend to converge to the same size. This yields the long-standing
problem of the bottleneck between condensation and
collision-coalescence.

The presence of broad droplet-size distributions has been detected via
\emph{in situ} measurements in warm clouds under very different
conditions, \cite{warner}. Still, this observation eludes full
theoretical understanding despite the number of different approaches
developed to this purpose. Some of them rely on the effects of
entrainment and mixing with non-cloudy air occurring in the regions of
the cloud near the boundaries (see, e.g., \cite{blyth}). However,
broad spectra are also observed inside the inner adiabatic cores, as
reported by \cite{brenguier}, where no boundary effects can possibly
explain their presence.  Droplets themselves have been proposed to be
sources of local variability in the environmental conditions. Indeed,
droplet evaporation is an internal source of turbulent kinetic energy
due to cooling associated to the absorption of latent heat, coupled
with buoyancy (see \cite{malinDNS,malinEXP98,malinEXP06}). Moreover,
the presence of droplets locally changes the water vapor content
through phase change, \cite{collinsReade}. The general conclusion is
that these microscopic fluctuations influence the process of mixing
occurring at the interface between cloud and clear-air. Also, the
presence of ultra-giant condensation nuclei has been proposed to
explain the large raindrops production (\cite{giant}). Other
properties such as salinity and surface curvature may produce absolute
and relative broadening, as proposed in \cite{korolevsalinity}.

Stochastic fluctuations of the environmental conditions, induced by
turbulence have been suggested as a broadening mechanism since the
'60s when the theory of stochastic condensation was first proposed,
see e.g. \cite{mazin,sedunov,bartlett}. This approach explains droplet
spectra broadening by observing that fluctuations of the ambient
conditions make droplets grow at different rates. This simple idea is
very powerful in that it concerns the whole cloud, regardless of the
presence of additional microscopic mechanisms, the influence of
ultra-giant nuclei and/or boundary effects such as mixing with dry air
and entrainment. Although all these ingredients play a role in the
huge natural laboratories represented by clouds, their contribution
may vary according to different conditions. On the contrary,
turbulence is a very good candidate as general broadening mechanism
within convective warm clouds. At a Reynolds number approaching $Re
\sim 10^8$, turbulence is known to be highly intermittent, with
statistics strongly deviating from a Gaussian one and a substantial
probability of fluctuations far exceeding the standard deviation (see,
e.g., \cite{frisch}). This means that droplets coming close one to the
other might have previously experienced disparate conditions, thus
invalidating any expectation based on uniform condensation. Here we
show that some qualitative features emerging from a simple idealized
setting may be well present in more complex models, just because they
are consequences of basic properties of turbulence itself. In view of
this purpose, under consistent hypothesis, we consider a simple but
rather general model for condensation to be detailed below.\\ 

Droplet growth in a fully developed turbulent flow can not be treated
analytically and the numerical analysis becomes a fundamental tool of
investigation. Numerical simulations of cloud turbulence are very
demanding because of the huge number of active degrees of freedom (the
microscale Reynolds number varies as a function of the spatial
resolution approximately as $Re_{\lambda} \sim N^{2/3}$). As a
consequence of the spatial structure of the turbulent fields, direct
numerical simulations (DNS) can either focus on large-scale structures
as done by \cite{epl05,jot}, or resolve the small scale features as in
\cite{VY01a,VY02b,korolevDNS,malinDNS}. The two choices are actually
mutually exclusive due to finite computing resources and both have
strong and weak points (see Section~\ref{sec:model}, for further
details).  Here we will try to take advantage of both these
approaches, to achieve a deeper understanding on the role of turbulent
fluctuations for the problem at hand.\\ In a nutshell, we first
evaluate the spreading of droplet-size distribution through a series
of direct numerical simulations, at increasing resolution, matching
the small scale features. Not surprisingly, each single DNS gives a
small degree of spreading, as already pointed out in previous works
focusing on small rising fluid parcels (see
e.g. \cite{VY02b,korolevDNS,malinDNS}). However this only tells us
that turbulence at the smallest scales alone fails to
explain the fast broadening of droplet-size distributions observed in
clouds. Then, we evaluate the dependence of the size spectra
broadening on the turbulent Reynolds number, i.e.~on the range of
spatial scales resolved. The broadening is found to increase with the
Reynolds number of turbulence.

Since the Reynolds numbers of real cloud turbulence are several orders
of magnitude larger than those described by DNS, this increase must be
accounted for when assessing the role of turbulence for condensation
in clouds. In order to get more insight on this point, we analyze the
trend of the size-spectra broadening as a function of the Reynolds
number. It is quite natural to derive, on the basis of consistent
assumptions, expectations for the observed broadening at large
Reynolds number. Within the model adopted, we derive upper and lower
bounds on the trend through a dimensional analysis of the system
equations. The upper bound is obtained by neglecting the vapor
depletion due to condensation onto cloud droplets. We argue that this
should be significant at small time lags and small Reynolds numbers,
and we show that this is consistent with two series of DNS
performed. We then reason that the lower bound, where the vapor
fluctuations are considered immediately depleted for
condensation/evaporation, should be more significant at high Reynolds
numbers and large time lags. A further DNS matching the large scales
of the problem suggests consistency with the argument.

The paper is organized as follows. In Section~\ref{sec:model} we
introduce the model for the time evolution of the vapor field and the
droplets advected by the turbulent air flow. Section~\ref{sec:DNS} is
devoted to the numerical approach and DNS details. Results concerning
the spreading of droplet-size spectrum are discussed in
Section~\ref{sec:results}, while in Section~\ref{sec:final}
conclusions and perspectives will follow.

\section{Model equations and numerical procedures}
\label{sec:model}
We focus on a turbulent velocity field advecting vapor and
droplets. The latter undergo size changes for evaporation or
condensation of the surrounding vapor. The three-dimensional velocity
field $\mathbf{v}$ evolves according to the Navier-Stokes equations
for an incompressible flow,
\begin{equation}
\label{eq:NS}
\partial_t \mathbf{v} +
\mathbf{v}\cdot\bm\nabla\mathbf{v}\,=\,- \frac{{\bm \nabla} p}{\rho_a} +
\nu\Delta\mathbf{v}+\mathbf{f}, \,\,\,\,\,\,\,\,\,\,\bm\nabla \cdot \mathbf{v}\,=\,0,
\end{equation}
\noindent
where $p$ is the pressure, $\rho_a$ is the air density and
$\mathbf{f}$ is an external statistically homogeneous and isotropic
forcing, providing a turbulent stationary flow. In eq.\ref{eq:NS},
$\nu \approx 0.15\, cm^2/s$ is the air kinematic viscosity. Note that
turbulence in clouds is generated by large-scale turbulent
fluctuations which can be strongly anisotropic. Indeed the presence of
gravity - here neglected - introduces a preferential direction through
large-scale thermal gradients and buoyancy, the engine of convective
motions.  However, we can assume that for sufficiently small internal
cloud cores, vertical stratification of the environment can be
neglected and the small-scale flow is essentially forced by nonlinear
transfer from larger scales, rather than by buoyancy. Isotropy can
thus be assumed for these scales of motion.  In Vaillancourt and Yau
(2000)\nocite{VY00}, the authors argue that this should be valid in
warm-cloud cores for spatial scales up to $L\sim 100\,m$. In addition
to large-scale thermal gradients, anisotropy is also produced through
buoyancy by microscopic temperature fluctuations. Sedimenting droplets
can evaporate, absorbing latent heat and thus locally cooling the
environment. There are conditions for which this effect can be
important for the cloud-clear air mixing at small turbulent kinetic
energy rates (\cite{malinEXP06,malinEXP98,malinDNS}). However,
provided that we focus on the inner adiabatic core, away from the
cloud boundaries, we can neglect this microscopic source of
anisotropic fluctuations as a first approximation. With
equation~(\ref{eq:NS}), we focus on the turbulent motion of in-cloud
air, neglecting the role of convective motions. Note that previous
two-dimensional DNS (\cite{epl05,jot}), suggest that the qualitative
effects of turbulence on condensation do not rely specifically on the
statistical details of the turbulent regime analyzed.

Water vapor molecules carried by the turbulent velocity field are the
source for droplet growth by condensation. The relevant quantity for
droplet condensation/evaporation is the supersaturation, which
quantifies the presence of vapor available for cloud particles
growth. Supersaturation is defined as $s := e/e_s-1$, where $e$ and
$e_s$ are the vapor pressure and the saturation vapor pressure
respectively. Droplets are able to grow when the surrounding vapor
content exceeds the saturation point ($s$ positive in
equation~(\ref{eq:R}), moist air). On the contrary droplets tend to
evaporate when $s$ is negative (dry regions).  Exhaustive \emph{in
  situ} measures of the small-scale statistics of the vapor field are
not available so far. Therefore, different models proposed cannot yet
be validated by direct comparison with real data.\\ For the sake of
generality, we adopt here the simplest generalization of the
well-known model proposed in Twomey (1959)\nocite{twomey}. While
Twomey considered a one-dimensional equation for the time-dependent
supersaturation function, here we consider the turbulent vapor to
fluctuate both in space and time. For this reason we introduce the
supersaturation as a field
$s(\mathbf{x},t):=\frac{e}{e_s}(\mathbf{x},t)-1$, that quantifies the
amount of vapor which is present at point $\mathbf{x}$ at time
$t$. Since $s(\mathbf{x},t)$ is allowed to fluctuate from positive to
negative values, dry and moist regions can coexist at the same
time.\\ The generalization of Twomey's equation (see also
\cite{prupp,shaw,korolevDNS}), for the supersaturation field is an
advection-diffusion equation:
\begin{equation}
\label{eq:s}
\partial_t s +\mathbf{v}\cdot\bm\nabla s \,=\, \kappa \Delta s + A_1 w -
\frac{s}{\tau_s}\,.
\end{equation}
Here we assume that the scalar field $s(\mathbf{x},t)$ is passively
advected by the turbulent flow $\mathbf{v}(\mathbf{x},t)$ which is not
affected by its presence. In this way, we are neglecting the
compositional effects of vapor on the buoyancy forces acting on the
flow, that are generally thought to be small (see \cite{stevens}). In
equation (\ref{eq:s}), $\kappa \approx 10^{-5} m^2 s^{-1}$ is the
molecular diffusivity of water vapor in air and $w(\mathbf{x},t)$ is
the vertical component of the turbulent velocity field
$\mathbf{v}$.\\ The term $A_1w$ acts as a source/sink term of
supersaturation resulting from the variation in temperature and
pressure with height. It can be computed by assuming 
that \emph{(i)} air pressure is quasi-hydrostatic, \emph{(ii)} the pressure 
of dry air is similar to that of moist air, \emph{(iii)} the motion 
inside the cloud core is adiabatic (temperature
changes with vertical motion according to a dry adiabatic lapse rate and
with latent heat effects as described below)
and \emph{(iv)} the atmosphere is not far from saturation. 
These standard arguments (see e.g.~\cite{prupp,korolevDNS}) 
yield $$A_1=\frac{Lg}{R_v c_p T^2}-\frac{g}{R_a T}$$ 
\noindent where $L$ is the latent heat of evaporation, $R_v$  and $R_a$ are the 
gas constants for vapor and dry air, $c_p$ is the specific heat at constant 
pressure, and $g$ is the gravitational acceleration. 
$A_1$ can be interpreted as a global supersaturation gradient.\\ 
The term $-s/\tau_s$ accounts for the double
effect of condensation/evaporation on supersaturation: on one side the
phase change directly modifies the water vapor content, on the other
side it locally modifies temperature due to absorption or release of
latent heat. In dry regions, $s(\mathbf{x},t) <0$, droplets tend to
evaporate releasing vapor and cooling the environment; here, the term
$-s/\tau_s$ is a source of supersaturation.  Conversely in moist
regions droplets tend to absorb vapor for condensation and to release
latent heat, and $-s/\tau_s$ acts as a sink term. The parameter
$\tau_s$ is the relaxation timescale of the supersaturation and
depends locally on the concentration and size of droplets. In each
elementary volume $V$, $\tau_s$ is defined as (see
appendix A):
\begin{equation}
\label{eq:taus}
\tau_s^{-1}=\frac{4 \pi \rho_w A_2 A_3}{V}\sum_{i=1}^n R_i\,,
\end{equation}
where $R_i$ are the radii of the $n$ droplets inside the considered
volume; $A_2$ is a function of thermodynamic parameters (see
appendix A); $\rho_w$ is the water density; $A_3$ is the
rate of droplet radius growth by condensation (see
equation~(\ref{eq:R})).  In the numerical procedure, we consider each
droplet to affect the value of $\tau_s$ in the eight nodes of the grid
cell surrounding its position. The weight of the contribution to each
node is calculated via a three-linear extrapolation.
Table~\ref{table:1} shows the reference values of these parameters
used in the numerical experiments.

Note that Twomey's one-dimensional equation for $s$ is derived under a
given temperature profile, neglecting fluctuations. It is difficult to
quantify temperature fluctuations in clouds, since \emph{in situ}
microscale cloud data are unavailable. For what concerns the vapor
field, this amounts to neglect an additional source of fluctuations
induced by temperature advection and diffusion.
The remarkable feature of the simple model identified by
equations~(\ref{eq:NS}) and (\ref{eq:s}) is that, despite its
simplicity, it allows to identify nontrivial mechanisms leading to the
spreading of the size spectra.

Given the evolution equation for the Eulerian turbulent fields, we can
now introduce the Lagrangian dynamics of cloud particles and the time
evolution of their radii. A complete description of the relation
between the water vapor and the size of a droplet would imply an
integral equation for the local dynamics occurring at the droplet
surface. Note that, however, the typical timescales associated with
the diffusional growth of an isolated droplet are much smaller than
the fastest timescale associated with turbulent changes in the ambient
conditions \cite{prupp}. Therefore we can consider the droplet to be
instantaneously in equilibrium with the surrounding vapor (a detailed
quantification of this assumption is given, e.g., in \cite{VY01a},
where the authors argue that it should be valid in the condensation
stage, for kinetic energy dissipation rates of about $\epsilon\approx
10^{-3}\,m^2s^{-3}$ that we consider here).  Under basic
assumptions\footnote{(i) Droplets are considered spherical
  (significant deformations from the spherical shape are typical of
  much larger drops, from sizes of hundreds of $\mu m$); (ii) the
  coefficient $A_3$ is not significantly altered by either the
  chemical composition of droplets or the size of the droplet itself
  (this holds for droplets already activated onto condensation
  nuclei); (iii) curvature and salinity corrections to the
  supersaturation at droplet surface (described by classical K\"ohler
  theory) are neglected for simplicity - these are generally thought
  to be small for radii larger than few microns (see
  e.g.~\cite{korolevDNS,prupp}).}  we end up with the following
equation for the i-th droplet growth rate (see also Pruppacher and
Klett (1997) for further details),
\begin{equation}
\label{eq:R}
\frac{dR_i(t)}{dt}=A_3\frac{s(\mathbf{X}_i(t),t)}{R_i(t)}\,.
\end{equation}
Here $A_3$ is a function of the local conditions, air temperature and
pressure, and assumed to be constant throughout the entire volume
(variations of this parameter with temperature in typical warm cloud
conditions are smaller than $3\%$).  According to
equation~(\ref{eq:R}), the growth rate varies from a droplet to
another, since it depends on the supersaturation fluctuation
$s(\mathbf{X}_i(t),t)$ measured along the trajectory $\mathbf{X}_i(t)$
of the single droplet. Due to turbulent transport, initially close
droplets separate very rapidly and eventually experience disparate
values of supersaturation throughout the entire cloud volume. This is
the most important difference with respect to Twomey's model where all
the droplets are exposed to the same supersaturation value.

Cloud droplets can be described as independent, Stokes particles,
whose trajectories $\mathbf{X}_i(t)$ and velocities $\mathbf{V}_i(t)$
evolve according to:
\begin{eqnarray}
\label{eq:X}
\frac{d\mathbf{X}_i(t)}{dt}\,&=&\,\mathbf{V}_i(t)\\[0.2cm]
\label{eq:V}
\frac{d\mathbf{V}_i(t)}{dt}\,&=&\,
-\frac{\mathbf{V}_i(t)-\mathbf{v}(\mathbf{X}_i(t),t)}{\tau_d^i}+g\hat{\mathbf{z}}.
\end{eqnarray}
\noindent
Here $\mathbf{v}(\mathbf{X}_i(t),t)$ is the fluid velocity at the
particle position; $\tau_d^i(t)=\frac{R_i^2(t)}{3\nu\beta}$ is the
particle response time (or Stokes time);
$\beta=3\rho_a/(\rho_a+2\rho_w)\approx 3\rho_a/(2\rho_w)$ is the
air/water density ratio; and $g$ is the gravity acceleration.
Equations~(\ref{eq:X}) and~(\ref{eq:V}), derived from the more general
treatment of Maxey and Riley (1983)\nocite{maxey}, are valid for
dilute suspensions of small spherical heavy particles. These
hypothesis are well verified during the condensation stage, as
discussed e.g. in Vaillancourt and Yau (2000). \\
Droplet statistics and dynamics in a turbulent medium can be studied
by focusing on possibly many different features, see e.g. Pinsky and
Khain (1997) for a review. Recently, also thanks to fundamental
improvements in the laboratory tracking techniques
(\cite{body,mordant,warhaft}), progress in the understanding of
particle motion in turbulence has undergone an explosive
development. Together with the experimental efforts, theoretical
analysis and numerical simulations have considered such different
issues as particle acceleration statistics (see
e.g.~\cite{warhaft,noiacc}), particle spatial inhomogeneous
distribution (see e.g.~\cite{aliseda,BFF01,CK04,clustering}),
collisional effects (see e.g.~\cite{WOKG05,wilkinson}). Granted that
in the core of a warm cloud all these features can be relevant, here
we focus on the particle size statistics during the early stages of
the condensational growth.

\section{Numerical simulations and range of parameters}
\label{sec:DNS}
As discussed earlier, DNS of cloud physics present a major problem:
there is a huge number of degrees of freedom that cannot be described
simultaneously. Turbulence is organized in spatial structures of
typical scales ranging from the large scale $L$ of hundreds of meters,
down to the Kolmogorov scale $\eta$ (typically $1\,mm$). Similarly,
the timescales range from thousands to fractions of a second. Within
this highly turbulent medium, a population of $10^{14}\div 10^{18}$
droplets evolve. Moreover, even if droplets are much smaller than any
turbulent eddy, their trajectory spans the whole range of turbulent
scales. This yields correlations with the fluctuations of the vapor
field - see \cite{epl05,jot} - and with the structures of the velocity
field as shown, e.g., in \cite{FE94}. Therefore, turbulent
motion at any scale plays a significant role in droplet
dynamics. However, when dealing with experiments {\it in silico},
because of computational limitations, it is compulsory to choose a
setting which describes only a limited range of fluctuations in the
system.
 
Recent results, reported in \cite{epl05,jot}, of direct numerical
simulations in two dimensions pointed out the importance of the
large-scale fluctuations of the vapor field. These provide a strongly
variable environment for droplet evolution resulting in a spreading of
the droplet-size spectrum. In such context, the small scales of
turbulence cannot be resolved and the analysis is limited to a
statistically representative subset of the whole population of
droplets. In \cite{VY02b}, the complementary setting is adopted: by
concentrating on a small rising parcel, the authors can consistently
describe the droplet evolution in full detail. This approach provides
small fluctuations that eventually produce a limited degree of
spreading. Note that the small scales inside a cloud are the endpoint
of a wide turbulent cascade involving a huge range of interacting
spatial and temporal scales. In particular, they are not independent
from large scales, suggesting that approaches which
separate small from large scales might fail in reproducing turbulent
effects. In this respect, since DNS do not describe the whole inertial range
of turbulence, the fluid-parcel approach does not represent a small
volume inside a huge cloud, but rather a very small cloud. \\

Here we wish to investigate how turbulence effects change with the
Reynolds number by merging the two complementary approaches above
mentioned. This is a crucial step in assessing the role of turbulence
for cloud-droplet condensation, since real cloud turbulence has
necessarily a dramatically higher Reynolds number than the simulated
flows. To this purpose, we perform two series of direct numerical simulations
at increasing resolution. The grid spacing $\Delta x$ of each
simulation corresponds to about $1\,mm$. By progressively increasing
the number of grid points for each spatial direction, we can resolve
larger integral scales $L$, defining the size of the cloud. More
precisely, we consider two series of four numerical experiments,
labeled as run $(a)$, $(b)$, $(c)$ and $(d)$, with $64^3$, $128^3$,
$256^3$ and $512^3$ grid points respectively. The two series of
simulations have different initial liquid water content (LWC):
$\approx\,1.2\,g/cm^3$ for series~1 and $\approx\,0.07\,g/cm^3$ for
series~2.  The integral scale of the system varies from $L\sim 9\,cm$
up to $L\sim 70\,cm$. The microscale Reynolds numbers $Re_{\lambda}
\approx \sqrt{15\,Re}$ (see \cite{frisch}) range from
$Re_{\lambda}\sim 40$ to $Re_{\lambda}\sim 185$, which represents the
state-of-the-art for DNS in cloud physics. \\The ratio between the air
kinematic viscosity and the vapor molecular diffusivity, also called
the Schmidt number, is $Sc=\nu/\kappa=1$, so that the flow and the
scalar dissipative scales are of the same order. Table~\ref{table:DNS}
lists all the relevant DNS parameters. Clearly the process of doubling
the resolution, if iterated, would ideally lead to the description of
the whole range of scales from $\eta \approx 1\,mm$ to $L \approx
100\, m$. As we can only perform the first few iterations of this
process, the next step will be to discuss if this can be extrapolated
to give information on Reynolds numbers which are not accessible by
DNS. An attempt in this direction will be discussed at the end of next
section with some detail.\\

\begin{figure}
  \begin{center}
    \includegraphics[height=7cm,draft=false]{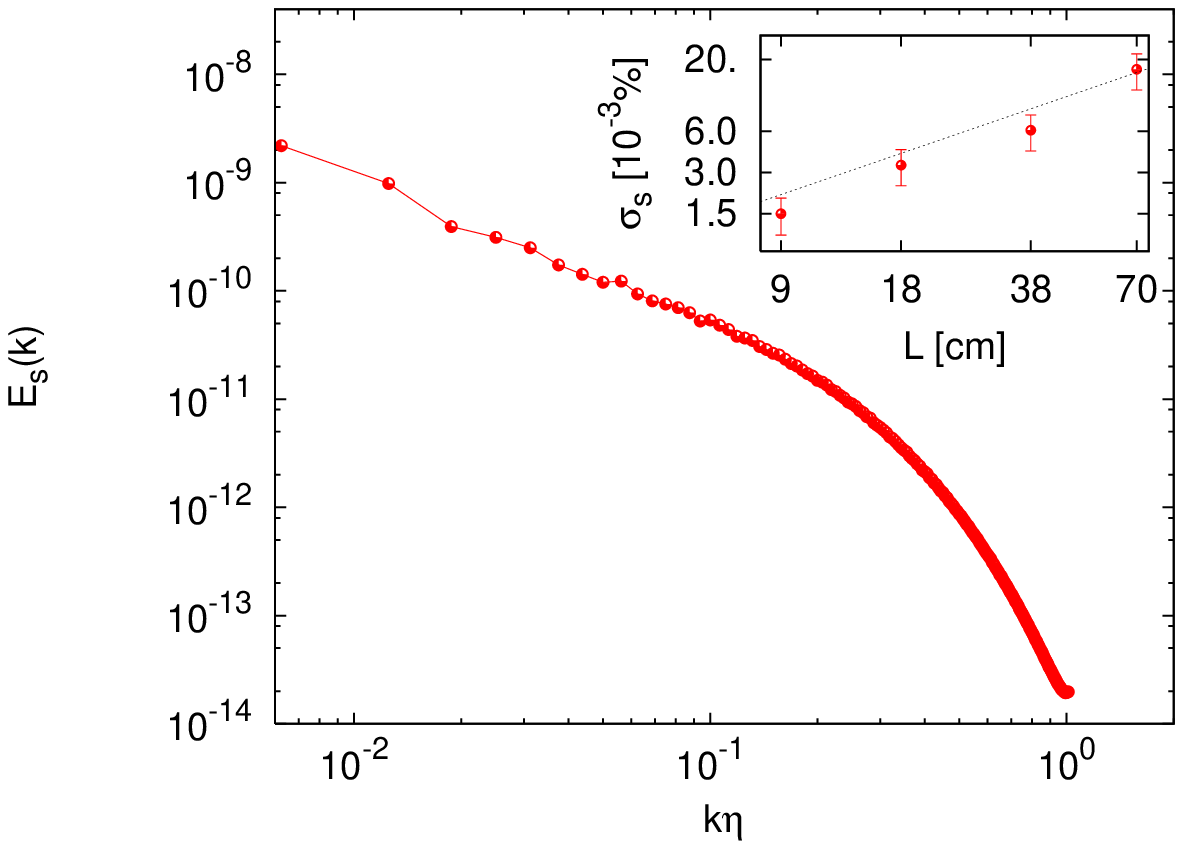}
  \end{center}
  \caption{Log-log plot of the stationary supersaturation spectrum for
    the run (d) at $Re_{\lambda}\sim 200$. It shows a $k^{-5/3}$ power
    law behavior, as expected from Kolmogorov-Obukhov-Corssin
    theory. The turbulent velocity field also displays a Kolmogorov
    spectrum (not shown). Inset: log-log plot of the standard
    deviation of the supersaturation field $\sigma^0_s$, measured in
    the stationary state, versus the size of the system $L$. The
    behavior is in agreement with the dimensional prediction
    $\sigma^0_s \sim A_1L$.}
  \label{fig:sspectra}
\end{figure}

For each DNS, equations (\ref{eq:NS}) and (\ref{eq:s}) are integrated
using pseudospectral methods with $2/3$-rule de-aliasing
(\cite{orszag}), in a three-periodic box. Molecular viscosity (and
diffusivity) is chosen so as to match the Kolmogorov lengthscale with
the grid spacing $\eta \simeq \delta x$: this choice ensures a good
resolution of the small-scale dynamics. Kinetic energy is injected at
an average rate $\epsilon$, by keeping constant the total energy in
each of the first two wavenumber shells (\cite{chen}). The scalar
field, also integrated on a triply periodic box, is forced by the
assigned gradient $A_1$; the term $-s/\tau_s$ does not contribute as
long as droplets are not injected into the flow. Time stepping is done
using a $2^{nd}$ order Adam-Bashfort scheme and the time step is
chosen to accurately resolve the smallest turbulent fluctuations and
the particle acceleration.\\ We obtain a statistically stationary
state for the velocity and supersaturation fields with no droplets by
integrating equations~(\ref{eq:NS}) and (\ref{eq:s}) for few
large-scale eddy turnover times $T_L=L/v_{rms}$.
Figure~\ref{fig:sspectra} shows the supersaturation spectrum at the
stationary state for run (d), $Re_{\lambda}\sim 200$, before particles
injection. In agreement with classical Kolmogorov-Obukhov-Corrsin
theory (see e.g. \cite{tennekes}), this exhibits a $k^{-5/3}$ power
law behavior in the Fourier space. Since the scalar spectrum is peaked
on the large scales, as the integral scale increases, we approach
larger and larger fluctuations. We indicate with $\sigma^0_s$ the
supersaturation standard deviation in the stationary state before
droplet injection. In the inset of Figure~\ref{fig:sspectra}, we show
that $\sigma^0_s$ increases linearly with the size of the system as
expected from a dimensional balance of terms in equation~(\ref{eq:s}),
yielding $\sigma^0_s \sim A_1 L$. Such increase can be understood from
the physical viewpoint: larger cloud sizes correspond to larger
displacements and stronger adiabatic cooling.  This directly provides
larger fluctuations in the vapor field through the term $A_1 w$ in
equation~(\ref{eq:s}). Without the feedback coming from droplets'
absorption of vapor, the value $\sigma^0_s$ would remain as in the
stationary state, i.e. $\sigma^0_s \sim A_1 L \sim A_1 w_{rms}T_L$.

Once the steady state has been attained, a monodispersed population of
droplets (with initial radius $R_i = 13 \, \mu m$ for series 1 and
$R_i = 5 \, \mu m$ for series 2) is injected into the flow. Droplet
concentration is for all experiments~$\approx130~cm^{-3}$, which means
that, on average, about one out of $2\,\div\,4$ cells contains a
droplet. In the largest simulation we followed the time evolution of
$32$ millions droplets. Initially, these are distributed randomly in
space, according to a statistically homogeneous Poisson distribution.
Lagrangian equations (\ref{eq:X}) and (\ref{eq:V}) for particle motion
are integrated simultaneously with those for Eulerian fields
(\ref{eq:NS}) and (\ref{eq:s}).  Particle initial velocities are set
equal to the local fluid velocity~\footnote{Tests have been performed
  by setting the droplet initial velocity equal to the terminal
  velocity $v_T=g \tau_d$. We did not observe any significant
  deviations in the results. Indeed particles rapidly equilibrate to
  the flow on a time scale of the order of their response time.}. To
obtain droplet velocity from equation (\ref{eq:V}), the underlying
flow velocity at the particle position has to be computed. This is
done via a linear interpolation in the three spatial directions
(\cite{yeung}), which was demonstrated to be adequate to obtain
well-resolved particle acceleration. Similarly, we compute the vapor
field $s(\mathbf{X}(t))$ at the particle position.\\
Coupling between droplets and the vapor field takes place via the term
$-s/\tau_s$ of eq.~(\ref{eq:s}), so that the initial vapor available
is consumed for condensation onto droplet surface or released for
droplet evaporation.  In typical cloud conditions, $\tau_s$ is of
order $1 \div 10\,s$, much smaller than typical large-scale eddy
turnover times.  As a result, the initial supersaturation standard
deviation is depleted from $\sigma^0_s\sim A_1 w_{rms} T_L$ to a lower
value, that can be estimated by a dimensional analysis of
equation~(\ref{eq:s}): $\sigma^\infty_s \sim A_1 w_{rms}
\tau_s$. 
We pointed out above that $\sigma^0_s$ grows with the size of the
cloud and the Reynolds number: note that $\sigma_s^\infty$ grows with
Reynolds as well, since we can estimate $w_{rms}\sim
v_{\eta}Re_\lambda^{1/2}$, where $v_{\eta}$ is the velocity
fluctuation at the viscous scale.  Considering that $\sigma^\infty_s
\sim \sigma^0_s \tau_s/T_L$, the ratio between the large-scale eddy
turnover time and the supersaturation absorption time is important to
evaluate the supersaturation fluctuations available for condensation
of cloud droplets. The complete system -flow, scalar and droplets- has
been studied, at the largest resolution, for about two large-scale
eddy turnover times. Longer time integrations were performed at lower
resolutions.

\section{Results and Discussion}
\label{sec:results}
\begin{figure}[h]
\begin{center}
\includegraphics[height=5.5cm,draft=false]{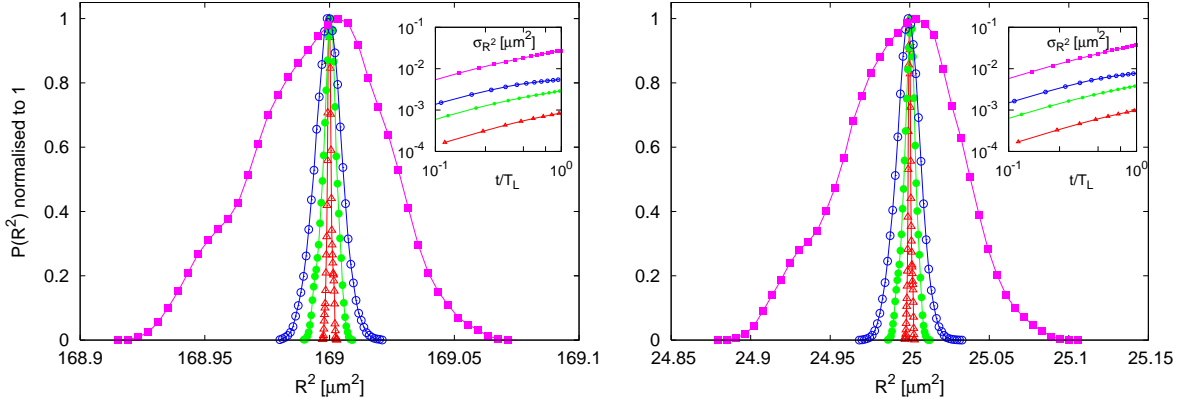}
\end{center}
\caption{Droplet square size distribution $P(R^2)$ measured after one
  large-scale eddy turnover time $T_L$, for the 4 DNS of series~1
  (left panel) and series~2 (right panel). At increasing the
  turbulence Reynolds number, we move from inner to outer
  curves. Symbols: run (a) red triangles; run (b) green dots; run
  (c) blue circles and run (d) purple squares. In each run,
  droplets' initial size distribution (not shown) is $\delta(R-R_0)$,
  with $R_0=13\, \mu m$ for series~1 and $R_0=5\, \mu m$ for
  series~2. Each simulation presents a small degree of spreading,
  which increases with the Reynolds number. Inset: log-log plot of the
  time evolution of the standard deviation of the square size
  distribution $\sigma_{R^2}$ for the runs (a)-(d) (from bottom to
  top). Symbols are the same of the main frame.}
\label{fig:PDFr}
\end{figure}

\begin{figure}
\begin{center}
\includegraphics[height=7cm,draft=false]{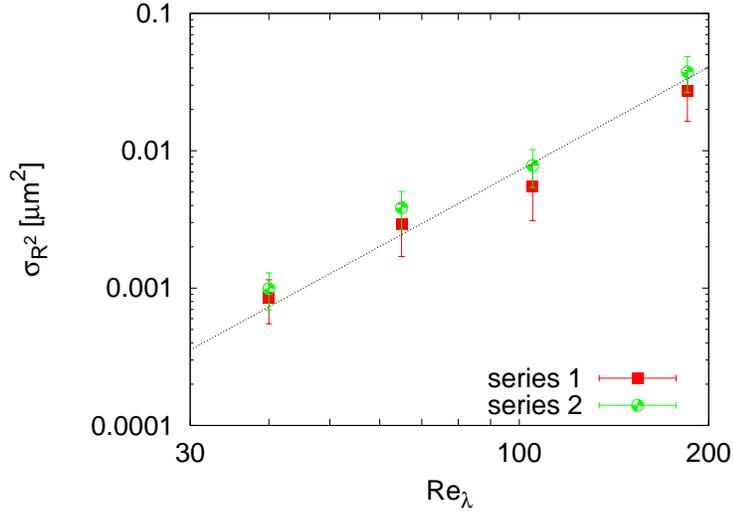}
\end{center}
\vspace{-0.1cm}
\caption{Log-log plot of the spreading of droplet size
  distribution $\sigma_{R^2}(T_L)$ for the square radius $R^2$,
  measured after one large-scale eddy turnover time $T_L$, as a
  function of the Reynolds number $Re_\lambda$. Data refer to
  simulations (a)-(d) of series~1 (red squares) and series~2 (green
  circles). For both series, the spreading is larger as the Reynolds
  number increases, since droplets evolve in conditions which are more
  and more differentiated. The dimensional prediction
  $\sigma_{R^2}\sim c_1 Re_{\lambda}^{\xi}$ with $\xi=5/2$, $c_1
  \propto A_1 A_3 v_\eta \tau_\eta^2$ is shown for comparison (see the
  text). The extrapolation of this law gives an upper bound to the
  final spreading for the target cloud of parameters $L=100\,m$,
  $\sigma^0_s \sim 2\,\%$ and $Re_{\lambda}\sim 4000 \div 7000$:
  $\sigma_{R^2}^{\it ext} \sim (200 \pm 100) \mu m^2$.  A correction
  of the expectation accounting for vapor depletion due to droplet
  feedback gives $\sigma_{R^2}^{\it ext} \sim (3.3 \pm 1.6)\,\mu m^2$
  for series~1, and $\sigma_{R^2}^{\it ext} \sim ( 9.3 \pm 4.5)\,\mu
  m^2$ for series~2 (see the text).}
\label{fig:trend}
\end{figure}

Our simulations start with a spatially uniform Poissonian distribution of
monodispersed droplets of radius $R_0=13\,\mu m$ for series~1 and
$R_0=5\,\mu m$ for series~2, and with vanishing acceleration. As the
cloud particles are released, they explore the entire volume and
experience the range of vapor fluctuations available in the system, of
initial standard deviation $\sigma^0_s$. During an initial transient
the fluctuations of the vapor field decrease due to the feedback of
droplets.\\ In Figure~\ref{fig:PDFr}, the droplet square size
distributions $P(R^2)$ are shown for runs (a)-(d) after one
large-scale eddy turnover time for series~1 and series~2. A small
degree of spreading is present for each simulation, increasing with
the size of the cloud~\footnote{Note that we impose a vanishing mean
  value for the initial supersaturation field, thus a constant mean
  droplet radius.}. This is due to the fact that - as already
discussed above - larger domain sizes correspond to larger
supersaturation fluctuations and longer large-scale eddy turnover
times. In other words, when evolving in a larger cloud volume,
droplets are exposed to more and more intense fluctuations of vapor,
for longer and longer times. The knowledge of these characteristic
times turns out to be crucial, as discussed in the sequel.\\ For $t
\leq T_L$, the standard deviation $\sigma_{R^2}$ of the square radii
$R^2(t)$ increases linearly in time as shown in the inset of
Figure~\ref{fig:PDFr}. This means that at short time lags droplet
surface grows with the vapor fluctuation initially experienced and
does not feel the underlying local variations~\footnote{A proportional
  linear growth is also observed for the standard deviation of droplet
  radii $R(t)$, (not shown). This is because, given the tiny
  supersaturation fluctuations, the droplet-size distribution is close
  to a Gaussian and the mean radius is much larger than the standard
  deviation. Hence $\sigma_{R^2}^2\approx
  2\sigma_R^4+4\sigma_R^2\langle R\rangle^2\approx 4\sigma_R^2\langle
  R\rangle^2$.}. After few absorption times $\tau_s$ have elapsed, the
vapor field standard deviation is depleted from $\sigma^0_s$ to
$\sigma^\infty_s$, thus slowing down the broadening of droplet size
distribution.  The mean vapor absorption time, estimated through
equation~(\ref{eq:taus}) with average concentration $\approx
130$~drops~$cm^{-3}$ and mean radius $13\,\mu m$ and $5\, \mu m$, turns
out to be $\approx 2.5\,s$ for series~1 and $\approx 7\,s$ for
series~2.  Since this time scale is comparable with the eddy turnover
time for simulations (a)-(d) (see table~\ref{table:1}
and~\ref{table:DNS}), the supersaturation fluctuations do not change
considerably from their initial value during $T_L$. For this reason,
the linear growth of $\sigma_{R^2}$ in time extends to the whole
large-scale eddy turnover time $T_L$ and the final spreading, measured
at $t=T_L$, is well approximated by:
\begin{equation}
\label{eq:dimensional}
\sigma_{R^2}(T_L)\sim 2A_3\sigma^0_s T_L\,.
\end{equation}
The square radius standard deviation $\sigma_{R^2}(T_L)$ is found to
be similar for the two series of simulations, pointing once more to a
limited role of vapor absorption by condensation when the Reynolds
number is small and $\tau_s\approx T_L$.\\ In Figure~\ref{fig:trend},
the standard deviation $\sigma_{R^2}(T_L)$ is shown as a function of
the microscale Reynolds number $Re_{\lambda}$ characterizing flows
(a), (b), (c) and (d).  We immediately observe that the final
broadening grows with the Reynolds number for both series of
simulations.  Given the following dimensional relations 
$T_L\sim \tau_\eta Re_\lambda$, $w_{rms}\sim v_\eta Re_\lambda^{1/2}$
and $\sigma_s^0 \sim A_1 w_{rms} T_L$, it is straightforward to
translate the simple expectation~(\ref{eq:dimensional}) in a scaling
relation with the Reynolds number:
\begin{equation}
\label{eq:scaling}
\sigma_{R^2} \sim A_3 A_1 v_\eta \tau_\eta^2 Re_\lambda^{5/2}\,,
\end{equation}
based on self-similarity of the growth process during the turbulent
regime. In Figure~\ref{fig:trend} we show for comparison the
dimensional expectation~(\ref{eq:scaling}).


 
It is natural to wonder whether this trend can give information on the
final broadening achieved at Reynolds numbers higher than
$Re_{\lambda}=200$.  This delicate point deserves insight since the
Reynolds number of real cloud turbulence typically achieve huge values
that cannot be described by direct numerical simulations.\\ At larger
Reynolds number and large-scale eddy turnover times, droplets
significantly absorb the surrounding vapor, so that $\sigma_s^0$ is no
longer a good approximation of $\sigma_s$. This means that the
scaling~(\ref{eq:scaling}) is an upper bound for the actual spectral
broadening at larger Reynolds number. A lower bound can be simply
obtained by replacing the supersaturation fluctuations with their long
time value $\sigma_s^\infty \sim \sigma_s^0 \tau_s/T_L$, yielding a
correction to the upper bound~(\ref{eq:scaling}) of a factor
$\tau_s/T_L$. Note that the scaling relation in~(\ref{eq:scaling})
would also change into a slower growth of the final broadening with
the Reynolds number, proportional to $Re_\lambda^{3/2}$.\\ The ratio
$\tau_s/T_L$ fluctuates both in space and time according to the local
properties of both turbulence and droplet population.  Given the
Kolmogorov timescale, we can estimate the large-scale eddy turnover
time at larger Reynolds numbers through $T_L\sim \tau_\eta
Re_\lambda$, as mentioned above. For a cloud of typical size $L\sim
100\,m$, liquid water content LWC$~1.2\, g/m^3$ with
$130$~drops~$cm^{-3}$ and Reynolds number\footnote{To put errorbars on
  the Reynolds number we consider fluctuations of $25\%$ on the
  average value, the same level of fluctuation observed in simulation
  runs (a)-(d).}  $Re_\lambda\sim (4000\div 7000)$, the large-scale
eddy turnover time is $T_L\approx 150\,s$ while $\tau_s$ does not
depend on Reynolds and its mean value is $\tau_s\approx 2.5\,s$, as in
series~1.  The extrapolation of the upper bound~(\ref{eq:scaling})
together with the lower bound gives: $ (3.3 \pm 1.6) \mu m^2
<\sigma_{R^2}(T_L) < (200 \pm 100) \mu m^2 $.

Validating these predictions by a direct numerical simulation,
describing the proper number of droplets and the whole range of
space-time scales involved, is of course not possible.  Instead, we
present a direct numerical simulation, labeled run~(e), designed so as
to reproduce the large-scale parameters typical of a cloud of size
$L=100\,m$: $T_L\approx 150\, s$, $\sigma^0_s\approx 2\%$,
$v_{rms}\approx 0.6~m s^{-1}$, $L\approx 100\,m$. The goal of this
further experiment is to verify that at long space and time scales
($T_L$ is now much larger than $\tau_s$), the lower bounds on vapor
fluctuations $\sigma_s$ and on the square radius fluctuations
$\sigma_{R^2}$ become more relevant. Of course, the small-scale
parameters do not match the realistic ones: in this run the smallest
resolved scales are $\eta \approx 25~cm$ and $\tau_\eta \approx 4\,s$,
and the turbulence Reynolds number is $Re_\lambda=185$. The initial
radius of droplets is $13\,\mu m$ and the numerical resolution is
$N^3=256^3$ grid points.  Space-time integration of the system has the
same features described in Section~\ref{sec:DNS}, and reference values
for the physical parameters entering the model equations are those
listed in Table 1.\\ Clearly, we can not follow the evolution of
$130\, cm^{-3}$ droplets, since in a volume of $(100)^3\,m^3$ they
would sum up to $N^*=1.3\times 10^{14}$.  The traditional cloud
physics approach avoids this problem by focusing on a number density
function representing the local concentration of droplets with a given
size (see e.g. \cite{malinDNS,jeffery}). However, a complete
continuous description for the turbulent transport of inertial
particles is still an open issue (see \cite{BCDLM07} and references
therein). 
Though more computationally
demanding, a Lagrangian approach turns out to be more appropriate. We
choose then to consider the complete evolution of a subset $N_{drops}$
of droplets, representative of the whole population.  Of course, the
whole population would absorb for condensation more vapor than the
representative subset does. Therefore an algorithm is needed to
account for the correct feedback of droplets on the supersaturation
field.  In order to accomplish this task, we simply normalize the
field $\tau_s$ with a factor $N^*/N_{drops}$.  Details on the meaning
and the convergence of this simple algorithm are discussed in
appendix~B.

\begin{figure}
\begin{center}
\includegraphics[height=7cm,draft=false]{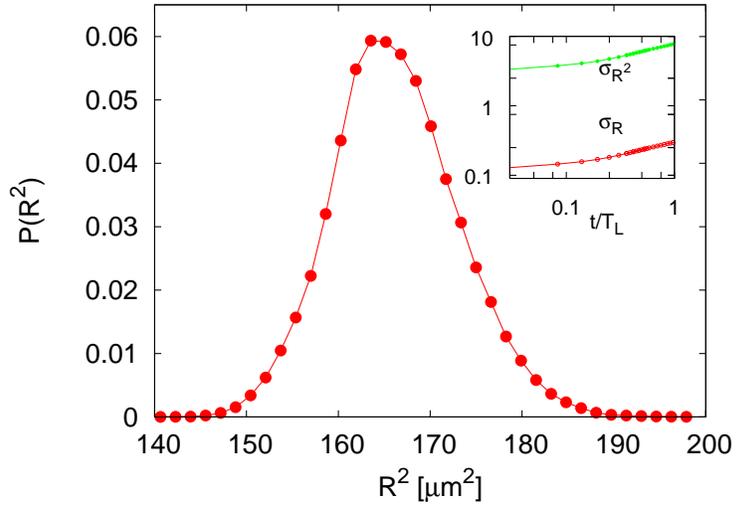}
\end{center}
\vspace{-0.1cm}
\caption{Droplet square size distribution $P(R^2)$ measured after one
  large-scale eddy turnover time $T_L\sim 150 \,s$ for run (e),
  matching the large scale cloud parameters. In the inset, time
  evolution of the standard deviation of the radius distribution,
  $\sigma_R(t)$, and of the square radius distribution,
  $\sigma_{R^2}(t)$. At time $t=T_L$, we measure $\sigma_R(T_L) \simeq
  (0.30 \pm 0.04)\, \mu m$ and $\sigma_{R^2}(T_L) \simeq (7 \pm
  1)\,\mu m^2$.}
\label{fig:PDFr-matchL}
\end{figure}

The supersaturation fluctuations for simulation (e) start with
$\sigma^0_s \sim A_1 w_{rms} T_L \approx 2\%$; after few vapor
absorption times they are depleted to $\sigma^\infty_s\approx 0.04\%$
and oscillate around this value for all later times.  The relevant
spreading of the square size distribution shown in
Figure~\ref{fig:PDFr-matchL} can be quantified in terms of the
standard deviation of the radius and of the square radius.  After one
large-scale eddy turnover time, we obtain $\sigma_R(T_L)\simeq(0.30\pm
0.04)~\mu m$ and $\sigma_{R^2}(T_L) \simeq (7 \pm 1)~\mu m^2$,
respectively. Although simulation (e) cannot resolve the whole range
of spatial scales of turbulence, by matching the parameters on the
large scales of the problem it is able to reproduce the intensity of
the large-scale fluctuations.

The final spreading achieved is very close to the lower bound,
obtained above by assuming that the supersaturation fluctuations
stabilize to $\sigma^\infty_s$ and rapidly forget their initial
condition.  This is consistent with the picture drawn and supports the
observation that the lower bound is more relevant when dealing with
large space-time scales. The result suggests that this could be the
case for large Reynolds number turbulence. Let us remark however that
care must be taken in this respect, since the spatio-temporal
complexity of turbulence at $Re_\lambda \approx 5000$ is not described
by the direct numerical simulation (e) presented.

\section{Conclusion and Perspectives}
\label{sec:final}
Turbulent fluctuations have been shown to play a role for the
broadening of the droplet-size distribution in an idealized setting
for condensation in warm adiabatic cloud cores.  A qualitative
explanation for this observation relies on the non-trivial spatial
structure of turbulent fields where droplets collect strongly
different histories.  The droplet size spectra broadening is shown to
depend on the actual range of spatial scales characterizing the
turbulent structure of the fields.  In particular, we show an increase
of broadening with the microscale Reynolds number by means of two
series of direct numerical simulations up to $Re_{\lambda}\sim 200$
and with millions of droplets.\\
The general outcome of this result is that the role of turbulence for
condensation in clouds may not be assessed just by focusing on small
Reynolds numbers, where a limited range of spatial scales is
considered.  Therefore, a strategy must be conceived to extrapolate
the results of DNS or laboratory experiments, that cannot achieve the
huge Reynolds numbers of real cloud turbulence.  This is also the case
for fluid parcel models which in fact do not represent a portion of a
large cloud, but rather a very small cloud.  Indeed, it is well-known
that initially close droplets inside a parcel would not remain close
for a long time, but separate explosively due to turbulent
transport.\\ In particular, though at the Reynolds numbers considered
the broadening is relatively small, it may potentially achieve
relevant values at the huge Reynolds numbers of real cloud
turbulence.\\ In order to get a better intuition on this potential, we
perform a dimensional analysis of the final broadening viewed as a
function of the Reynolds number.  At moderate Reynolds numbers, it is
possible to neglect the feedback of droplets on the vapor field. This
yields a dimensional expectation consistent with the DNS data. \\ At
larger Reynolds numbers, however, droplets are expected to
consistently modify the surrounding vapor field.  Therefore, the
extrapolation of the dimensional behavior provides an upper bound to
the actual spectral broadening at large Reynolds numbers.  A
correction is proposed to estimate the effective absorption of vapor
for condensation at large Reynolds numbers, yielding a lower bound on
the spectral broadening.  A further simulation designed to discuss the
validity of this argument shows that at large space-time scales the
behavior is well approximated by the lower bound. This suggests,
within the limits of simple dimensional arguments, that the lower
bound could be a good candidate to estimate the size-spectra
broadening at large Reynolds numbers.

In this work, we focused on the role of turbulence for droplet
condensation at moderate-high Reynolds numbers.  In order to enlighten
the basic mechanisms yielded by turbulence, we considered a simple
model describing the essential physical processes.  More sophisticated
models have been proposed in the literature accounting for additional
microphysical and thermodynamic couplings.  Ingredients such as the
explicit dependence of the vapor field on the temperature
fluctuations, the microscopic interactions of droplets with the
turbulent fields and the role of buoyancy both at small and large
scales may yield corrections to the presented results.  Other aspects,
such as droplet preferential concentration and modified relative
velocity effects, were not specifically addressed in this work, though
they are in fact included in the model. These have been associated to
the efficiency of collision/coalescence processes (see
\cite{WOKG05,wilkinson}), before they become dominated by
gravity. \\
The problem here considered clearly deserves further theoretical,
experimental and numerical insight and represents a promising
challenge for future research.  In particular, the ideas at the basis
of this work could be well explored through laboratory
experiments. The higher Reynolds numbers covered would allow to verify
and complement the picture drawn through the numerical analysis
presented.

\begin{acknowledgment}
  We acknowledge discussions with Antonio Celani and Andrea Mazzino,
  who have inspired and motivated this work. AS was partially
  supported by L'Or\'eal Italia - Unesco For Women in Science
  Fellowship, and by HPC-Europa Transnational Acces Program. AL
  acknowledges discussions with Andrea Buzzi and support from CNR
  grant ``Short-Term Mobility''. Numerical simulations were performed
  at the supercomputing center CINECA (Italy). Raw data from our
  numerical simulations can be downloaded freely from the web site of
  the iCFDdatabase {\sc http://cfd.cineca.it}.
\end{acknowledgment}

\begin{appendixA}
\appendix{\begin{center}Water-vapor interaction\end{center}}
\label{appendixA}
The expression of the absorption time $\tau_s$ can be computed
directly from the classical form of Twomey's model by identifying the
terms:
\begin{equation}
\label{eq:sovertaus}
\frac{s}{\tau_s}=A_2\frac{d\rho_L}{dt}, \quad\quad\quad\quad\quad \textrm{where }\quad
A_2=\frac{R_aT}{\varepsilon_{wa} e_s}+\frac{\varepsilon_{wa} {\cal L}^2}{p T c_{pa}}
\end{equation}
\noindent
where $\rho_L=m_w/V$ is the density of liquid water of mass $m_w$
inside volume $V$; $\varepsilon_{wa}$ is the ratio between the
molecular weight of water and dry air; ${\cal L}$ is the latent heat
of water evaporation; $R_a$ is the specific gas constant for dry air;
$T$ is the absolute temperature and $c_{pa}$ is the specific heat of
dry air at constant pressure $p$ (see Pruppacher and Klett
(1997)). The evaluation of the constant $A_2$ in typical warm cloud
conditions gives the reference value shown in Table~\ref{table:1}. Its
fluctuations with temperature are less then $1\%$, so that we assume
$A_2$ to be constant.\\ Since we focus on inner cloud cores, we
neglect entrainment; then droplets inside the volume $V$ are the only
responsible for the local change in vapor content:
\begin{equation}
\label{eq:ricavoTaus}
\frac{d\rho_L}{dt}=\frac{1}{V}\frac{dm_w}{dt}=\frac{1}{V}\sum_{i=1}^n4\pi\rho_w 
R_i^2\frac{dR_i}{dt}=\frac{4\pi\rho_wA_3}{V}\sum_{i=1}^nR_i \,s\,,
\end{equation}
where (see eq. (13-28) of Pruppacher and Klett (1997), ignoring small curvature and solute terms for activated droplets),
\begin{equation}
\label{eq:A3}
A_3=\left(\frac{R_vT}{\rho_w D_v e_s} + \frac{{\cal L}^2}{\rho_w k_a
  R_vT^2}\right)^{-1}\,,
\end{equation}
\noindent
where $R_v$ is the specific gas constant for moist air, $\rho_w$ is the density of water,
$D_v$ is the thermal diffusivity of water vapor in air, and $k_a$ is the air
thermal conductivity.\\ In eq.~(\ref{eq:ricavoTaus}), $R_i$ are the radii
of the $n$ droplets inside the volume $V$.  The rate of variation of
the radius is given by equation~(\ref{eq:R}), where $s$ is considered
equal for each droplet inside the small volume $V$.  In the numerical
analysis, $V$ is a cube of edge $\eta$, similar to the vapor diffusive
scale.  The fluctuations of the scalar field are tiny under this scale
and all the droplets inside the volume $V$ experience approximately
the same value of $s$. In this sense, the Twomey's parcel, where no
spatial fluctuations of the supersaturation field are accounted for,
corresponds to our grid cell.\\From equations~(\ref{eq:sovertaus}) and
(\ref{eq:ricavoTaus}) we end up with the expression~(\ref{eq:taus})
for the absorption time $\tau_s$.
\end{appendixA}
\newpage

\begin{figure}
\includegraphics[height=5.5cm]{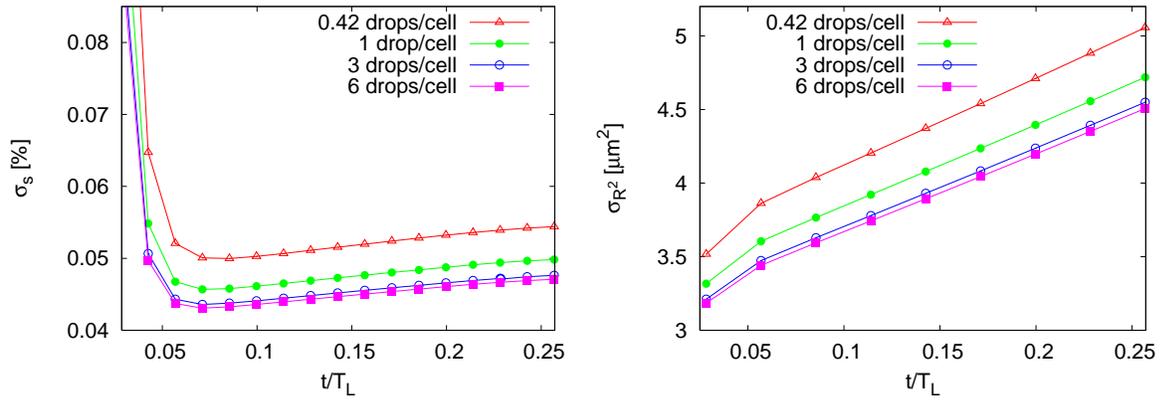}
\caption{Left panel: Supersaturation standard deviation as a function
  of time. Right panel: Droplet-square-radius standard deviation
  $\sigma_{R^2}$ as a function of time.  
  Results are shown for simulation~(e), with
  $4$ different numbers of droplets $N_{drops}$ representative of a
  population of $N^*\approx 10^{14}$ droplets. The feedback of the
  whole population on the supersaturation field is accounted for by
  renormalizing $\tau_s$ with a factor $N^*/N_{drops}$ (see text).
  The algorithm converges when the number of representative droplets
  is larger than $1/cell$.}
\label{fig:B1}
\end{figure}

\begin{appendixB}
\appendix{\begin{center}Renormalization of droplets population\end{center}}
\label{appendixB}

Simulation (e) is designed to describe a volume of $(100\,m)^3$ with
$130\,cm^{-3}$ droplets.  A major drawback of this simulation is that
it can not describe turbulence structure at the smallest scales, below
$\eta=25 \,cm$ and $\tau_{\eta}=4\,s$. Also it can not follow the
individual history of the correct number of droplets. This might
influence the statistics of vapor fluctuations since the correct
feedback of droplets on the vapor field is not accounted for.  In
particular, a concentration of $130\,cm^{-3}$ droplets yields
$N^*\approx 10^{14}$ total number of particles, whereas we consider
the evolution of several millions droplets ($N_{drops}$) due to
computational limits. Clearly, $10^{14}$ droplets would absorb much
more vapor for condensation than a few millions do, thus an algorithm
is needed in order to account for this. The simplest way to estimate
the feedback of the whole population on the supersaturation field is
to normalize $\tau_s$ with a factor $N_{drops}/N^*$.  This amounts to
consider each droplet as representative of $N^*/N_{drops}$
\emph{equal} particles in the same volume $(\delta x)^3$, where
$\delta x \approx 25 \,cm$.\\ To evaluate the reliability of this
approach, we must consider the single grid cell. If there is $1$
representative drop per cell, the algorithm computes the feedback on
vapor as if the average size of droplets in the cell was exactly the
radius of that unique representative.  On the contrary, a volume of
$(25\,cm)^3$ - the grid cell size in simulation (e)-, should contain
several droplets spanning a whole size spectrum whose mean value is
not well represented by one single droplet. Of course, a higher number
of particles would better represent the local mean radius and we
expect the algorithm to converge for density values larger than $1$
droplet per cell. We test this expectation, by repeating simulation
(e) for a time lag $t=T_L/4$, with $0.4$, $1$, $3$ and $6$ droplets
per grid cell, which correspond to a total number of $N_{drops}
\approx 7$, $17$, $50$ and $100$ millions droplets,
respectively. Consistently, we use different renormalization factors
$N^*/N_{drops}$. On average, the supersaturation absorption timescale
$\tau_s$ is the same in the four simulations (not shown), pointing to
a correct renormalization.

In figure~\ref{fig:B1}, left panel, the four standard deviations of
the supersaturation field are shown in time. They all start with the
same initial value (about $A_1 w_{rms} T_L \approx 2\%$ - not shown),
and relax to a value of the order of $A_1 w_{rms} \tau_s$, with some
differences for the different concentrations of representative
droplets. These deviations decrease when the number of representative
droplets increases, showing that the algorithm renormalizing the
feedback of droplets on vapor converges. Consistently, the evolution
of the size spectra broadening (shown in figure~\ref{fig:B1}, right
panel) tends to collapse to a unique curve when the number of droplets
per cell is larger than $1$. According to this analysis, we choose to
perform the complete simulation lasting one large-scale eddy turnover
time with $1$ and $3$~droplets per cell, which correspond to a total
number of about $17$ and $50$ millions droplets, respectively.  The
error bar on the final broadening is estimated from these two runs.
\end{appendixB}

\clearpage
\bibliographystyle{ametsoc}
\bibliography{references}

\clearpage

\clearpage
\begin{table}
  \caption{\label{table:1} Reference values for the physical
    parameters used in the numerical experiments. $A_1$ is the global
    supersaturation gradient. $A_2$ and $A_3$ are functions of the
    ambient thermodinamic parameters (see appendix A): as reference
    values for the temperature and pressure we used $T=283 K$ and $p
    \approx 1000 hPa$. The values of the vapor relaxation time
    $\tau_s$, the droplet radius $R$ and the Stokes number $St$
    correspond to averages on the initial condition of droplet
    population.}

  \begin{center} 
    \begin{tabular}{ccccccccc}
      \hline $label$ & $A_1$ & $A_2$& $A3$ & $N_{drops}/V$ & $LWC$ & $\tau_s$ &
      $R$ & $St$\\ $ $ & $(m^{-1})$ &$(kg^{-1}m^3)$ & $(\mu\,m^2s^{-1})$ &
      $(cm^{-3})$ & $(g/m^3)$ & $(s)$ & $(\mu\,m)$ & $ $\\
      \vspace{0.1cm}\\
      \hline 
      \hline
      \vspace{0.2cm}
      $series~1$ & $5 \times 10^{-4}$ & $350$ & $50$ & $130$ & $1.2$ & $2.5$ & $13$ & $3.5\times 10^{-2}$ \\
      $series~2$ & $5 \times 10^{-4}$ & $350$ & $50$ & $130$ & $0.07$ & $7$ & $5$ & $5\times 10^{-3}$ \\
      \hline
    \end{tabular}
  \end{center}
\end{table}

\begin{table} 
  \caption{\label{table:DNS}Parameter of the DNS, series 1 and 2. From
    left to right: number of gridpoints $N^3$, integral scale $L$,
    large-scale eddy turnover time $T_L$, microscale Reynolds number
    $Re_\lambda$, average kinetic energy dissipation rate $\epsilon$,
    Kolmogorov spatial scale $\eta$, Kolmogorov timescale $\tau_\eta$,
    initial supersaturation standard deviation $\sigma^0_s$, velocity
    standard deviation $v_{rms}$ and number of droplets $N_{drops}$.}
  \vspace{0.3cm}
  \begin{tabular}{ccccccccccc}
    \hline
    label & $N^3$ & $L$ & $T_L$ & $Re_{\lambda}$ & $\epsilon$ & $\eta$ & $\tau_{\eta}$ & $\sigma^0_s$ & $v_{rms}$ & $N_{drops}$\\
    &   & $(cm)$ & $(s)$ &  & $(m^2s^{-3})$ & $(cm)$ & $(s)$ &$(\%)$  & $(ms^{-1})$ & ($\times 10^5$)\\
    \hline
    \hline
    \vspace{0.1cm}\\  
    (a) & $64^3$ &  $9$  & $2.0$ & $40$ & $10^{-3}$ & $0.1$ & $0.1$ & $1.5 \times 10^{-3}$ & $4\times10^{-2}$ & $0.93$ \\ 
    (b) & $128^3$ & $18$ & $3.5$ & $65$ & $9.\times 10^{-4}$ & $0.1$ & $0.1$ & $3.4 \times {10^{-3}}$ & $5.\times 10^{-2}$ & $8.2$\\ 
    (c) & $256^3$ & $38$ & $5.5$ & $105$ & $10^{-3}$ & $0.1$ & $0.1$ & $6.1 \times 10^{-3}$ & $7.\times 10^{-2}$ & $71.2$\\ 
    (d) & $512^3$ & $70$ & $7.6$ & $185$ & $1.1 \times 10^{-3}$ & $0.1$ & $0.1$ & $1.2 \times 10^{-2}$ & $1.\times 10^{-1}$ & $320$\\ 
    \hline
  \end{tabular}
  \vspace{0.2cm}
\end{table}

\end{document}